\documentstyle[version2,aps,preprint]{revtex}
\date{November 3, 1993}

\begin{document}
\draft
%\preprint{\small UBCTP-92-28, CTP\#2157}
\begin{title}
{\bf Cooper pair tunnelling into a}
{\bf Quantum Hall fluid}
\end{title}
\author{Matthew P.A. Fisher }
\begin{instit}
Institute for Theoretical Physics\\
University of California\\
Santa Barbara, CA   93106-4030
\end{instit}

\begin{abstract}
Transport through a tunnel junction connecting a superconductor to a
spin-aligned quantum Hall fluid at
filling $\nu$ is studied theoretically.  The dominant transport channel at
low temperatures is the tunnelling of Cooper pairs into edge states of the
quantum Hall fluid.  This process, which is
greatly suppressed at low energies due to both
Coulomb and Pauli exclusion effects, leads to
a tunnelling conductance which vanishes with temperature as $T^{4/\nu -2}$,
for $\nu^{-1}$ an odd integer.  For integer fillings with $\nu >1$ the "Pauli
blockade" is circumvented and
a non-vanishing conductance is predicted.
\end{abstract}
\pacs{PACS numbers: 73.40.Gk   74.50.+r}
\narrowtext

There are some striking parallels between the phenomena
of the fractional quantum Hall effect\cite{Girvin} and superconductivity.  In
both cases
one has a system which exhibits dissipationless flow of electrical current.
In each case, the physics was elucidated initially by a many-body
wave function, involving a new charge carrier.  More recently a Ginzburg-Landau
approach to
the quantum Hall effect has identified an underlying condensed
boson\cite{Zhang},
responsible for the dissipationless flow, highlighting the analogy between
the two phenomena.
Very recently
attention has shifted to the study of weak links or point contacts in the
quantum Hall effect\cite{weak}, the loose analog of a Josephson junction.

Since 1992 several experimental groups\cite{exp1,exp2,exp3} have succeeded in
making
low resistance contacts between superconducting
leads and a two-dimensional
electron gas (2DEG) in a semiconductor heterostructure.  Evidence has been
found for Andreev reflection\cite{Andreev}, in which Cooper pairs
are converted into unpaired electrons and transported into
the 2DEG.  These experiments raise the exciting possibility of
making and studying a tunnel junction between a superconductor and a
dissipationless quantum Hall fluid.  Motivated by this,
I consider in this paper
the simplest possible tunnelling model for a such a tunnel junction.  I focus
exclusively on
low temperature d.c. electrical transport through the junction, which
is dominated by pair tunnelling.

An immediate obstacle arises when considering a tunnel junction between a
superconductor and a Hall fluid.  The large magnetic field
needed to put the 2DEG into the quantum Hall fluid will tend to suppress
the superconductivity.  One is thus restricted to type-II superconductors,
with upper critical fields $H_{c2}$ in the 5-10 Tesla range or above.
In this case, magnetic vortices will penetrate the superconductor,
but provided they are strongly pinned (ie in the vortex-glass
phase\cite{vortex})
they will not contribute to the low temperature transport of interest.

In addition to penetrating the superconductor, the large magnetic field will of
course tend to spin polarize the 2DEG.  Here I consider only the case of a
fully spin polarized electron gas.  Then a spin singlet Cooper pair passing out
of the superconductor into the 2DEG must not only break apart into two unpaired
electrons, as in a conventional Andreev process\cite{Andreev}, but must undergo
a spin-flip process to align the spin of both electrons with the
magnetic field.  The suppression of the tunnel current due to
this spin-flip process
can be studied independently of the quantum Hall effect by
placing the external magnetic field in the plane of the 2DEG.  Of interest
here, though, is when the field is perpundicular, and the 2DEG is in a
dissipationless
quantum Hall fluid.  Then the
transport current in the 2DEG will be confined to edge
states\cite{edge1,edge2}.  Consequently,
at low temperatures the dominant
transport process through the junction should be tunnelling of a Cooper pair
into the edge state.

In this paper I introduce and analyze a simple model for Cooper pairs
tunnelling into a quantum Hall edge state.  I focus primarily on odd integer
$\nu^{-1}$, where the edge state is believed to be a single channel chiral
Luttinger liquid\cite{edge2}.  As recently emphasized\cite{Kane}, the
tunnelling of electrons into a Luttinger liquid is
greatly suppressed at low energies, and leads to a true Coulomb blockade
with vanishing tunnelling conductance at $T=0$.  Here,
though, $\it{two\/}$ electrons (a Cooper pair) are trying to tunnel
$\it{simultaneously\/}$ into the Luttinger liquid edge state.
The central point of this paper, established below, is that the Pauli exclusion
principle operating between the $\it{pair}$ of
electrons, leads to an additional suppression of the tunnelling
conductance, over and above the Coulomb blockade.  Together, the
Coulomb blockade and this new "Pauli blockade" effect,
lead to a conductance which is predicted to vanish with a large power
of temperature:
$G \sim T^{4/\nu -2}$, for $\nu^{-1}$ an odd integer.  Strikingly, the
predicted conductance
vanishes even for $\nu=1$, when the edge state is a Fermi
liquid\cite{edge1,edge2}.
In this case, the vanishing conductance is due entirely to
a Pauli exclusion between the pair of electrons.  For integer filling with
$\nu>1$, where the edge consisits of
more than one Fermi liquid channel, the Pauli blockade is circumvented, and
a non-vanishing low temperature tunnelling conductance is predicted.

It is perhaps surprising that for the fractional Hall state at filling
$\nu=1/3$ the tunnelling conductance is predicted to
vanish with such an enormous power of temperature, $G \sim T^{10}$.  Despite
the
fact that both the superconductor and quantum Hall fluid have
zero resisitance, the junction between the two is predicted to be an extremely
good $\it{insulator}$.  Ultimately, this is because very different bosons are
condensing in the two
systems, a Cooper pair in one and a vortex-electron composite\cite{Zhang} in
the other.

Consider then a tunnel junction, or point contact, between a singlet
superconductor and
an incompressible quantum Hall fluid at filling $\nu$, as depicted
schematically in the Figure.  The total
Hamiltonian is expressed as a sum of three pieces:  $H=H_{QHE} + H_{sc} +
H_{pert}$,
where $H_{QHE}$ is the Hamiltonian for the spin-polarized quantum Hall fluid,
$H_{sc}$ is the Hamiltonian
for the superconductor and $H_{pert}$ a perturbation which couples
them together.  Being interested only in temperatures well
below the superconducting gap, it is adequate to model the superconductor in
terms of a bosonic pair field, $\hat{c}$,  which exhibits
long-ranged (vortex-glass\cite{vortex}) order and has a non-zero condensate,
$<\hat{c}>=\Delta$.
Moreover, for temperature scales well below the gap in the
quantum Hall fluid, $H_{QHE}$ can be taken as an edge Hamiltonian (see below).
For the perturbation term at the tunnel junction
we take initially:

\begin{equation}
H_{pert} =  \int_{x,x^\prime} t_1(x,x^\prime) [\psi^\dagger_\uparrow (x
) \psi^\dagger_\downarrow
(x^\prime) \hat{c}(x=0) + h.c.] + \int_x t_2(x) [\psi^\dagger_\uparrow (x)
\psi_\downarrow (x) + h.c.]  ,
\label{pert}
\end{equation}
where $x$ is a 1d spatial coordinate which runs along the edge of the quantum
Hall fluid.
The first term hops a Cooper pair from the superconductor through the point
contact (at $x=0$) into the edge of the Hall fluid.  Since the Cooper pair is
a singlet, the two electrons deposited are of opposite spin.
The pair "wave function", $t_1(x,x^\prime)$, is symmetric under interchange
of $x$ with $x^\prime$.  It is assumed to fall off exponentially for $x$ and
$x^\prime$ large compared
to the pair size - essentially the superconducting coherence length $\xi$.
Since the 2DEG is
completely spin polarized by the magnetic field (spin up), the edge state only
transports spin up electrons.  It is thus necessary to consider spin-flip
processes.  We model these
phenomenlogically by the second term in (1), which flips the electron spin
with amplitude $t_2$ at position x along the edge.  Physically,
$t_2$ will probably be dominated by spin-orbit mediated scattering
in the 2DEG. (It is also possible that nearby magnetic impurities mediate the
spin flip process.)

It is apparent from $H_{pert}$ in Eqn. (1), that at second order
in $t_j$ a term will be generated which destroys a Cooper pair in the
superconductor
and creates two up spin electrons in the 2DEG.  Retaining only this composite
process, we let $H_{pert} \rightarrow H_{tunn}$ with

\begin{equation}
H_{tunn} = \int_{x,x^\prime} t(x,x^\prime) [\psi^\dagger_\uparrow (x)
\psi^\dagger_\uparrow
(x^\prime) \hat{c}(x=0) + h.c.] .
\label{tunn1}
\end{equation}
where $t(x,x^\prime) = t_1(x,x^\prime) [t_2(x)-t_2(x^\prime)]$.  Since
$t_1(x,x^\prime)$ vanishes rapidly for $x,x^{\prime}$ larger than the coherence
length $\xi$, the integrals above will be dominated by small x and $x^\prime$.
For simplicity, we replace these integrals by a single term,

\begin{equation}
H_{tunn} = t [\psi^\dagger_\uparrow (x=\xi) \psi^\dagger_\uparrow
(x=0) \hat{c}(x=0) + h.c.] .
\label{tunn2}
\end{equation}
where the spatial arguments are separated by the coherence length, $\xi$. The
qualitative results
obtained below do not depend on this simplification.  The remaining parameter,
$t$, characterizes the strength of the pair tunnelling.  In the following
we will drop the spin subscript on the up-spin electrons.

For the integer quantum Hall state, with $\nu=n$, the
edge states, which carry away the spin-polarized
electrons, are non-interacting Fermi liquids\cite{edge1}.
The appropriate effective (Euclidian) action in this case is simply

\begin{equation}
S_{QHE} = \sum_{j=1}^{n} \int dx d\tau \psi^\dagger_j (\partial_\tau
-iv_j \partial_x ) \psi_j
\label{action1}
\end{equation}
where $\tau$ is imaginary time.  Here $\psi_j$ denotes the (spin-up) electron
in edge branch j and  $v_j$ is the corresponding edge velocity.  For fractional
states at odd integer $\nu^{-1}$,
the edge state is expected to be a single-channel chiral Luttinger liquid.  The
appropriate
Euclidian action in terms of a chiral boson field $\phi$ is\cite{edge2}:

\begin{equation}
S_{QHE} = {1 \over {4\pi \nu}} \int dx d\tau
\partial_x \phi [i \partial_\tau \phi + v \partial_x \phi ]  ,
\label{action2}
\end{equation}
with $\nu^{-1}$ an odd integer.  The electron operator is given by $\psi \sim
e^{i\phi/\nu}$.

Consider now the effect of the tunnelling term Eqn. (3) in transferring charge
across the junction.  Our approach is perturbative in the
tunnelling amplitude $t$.  Before calculating the tunnelling conductance, it is
instructive to consider a simple renormalization group (RG)
transformation\cite{Kane}
which tells us how the tunnelling amplitude $t$ varies with the energy (or
temperature) scale.
Since $H_{tunn}$ in Eqn.(3) involves fields near $x=0$, it is
useful to integrate out the degrees of freedom for
$x\neq 0$ in both the edge action above, and in the superconductor.
Since the superconductor has a non-zero condensate, it is
legitimate to simply replace the operator $\hat{c}$ in Eqn. (3) by the c-number
$\Delta$.  The remaining field is the spin-up electron near $x=0$,
which depends on imaginary time $\tau$.
Consider an RG transformation which integrates out a
shell of Matsubara frequencies between $\Lambda/b$ and $\Lambda$,
where $\Lambda$  is a high frequency cutoff.  The resulting RG flow equation
for the tunnelling amplitude $t$
is given to leading order by ($l=e^b$):

\begin{equation}
\partial t/ \partial l = (1-\delta)t  ,
\label{flow}
\end{equation}
where $\delta$ is the scaling dimension of the tunnelling operator,
$O = \psi^\dagger (x=\xi) \psi^\dagger (x=0)$.  This dimension
can be evaluated
from the large (imaginary) time decay of the correlation function:

\begin{equation}
<O^\dagger (\tau) O(\tau=0)> \sim \tau^{-2\delta}   ,
\label{dim}
\end{equation}
using the edge action in Eqn.(4) or (5).

Specializing to odd integer $\nu^{-1}$, the tunnelling operator can be
expressed in terms of the boson field $\phi$, as $O \sim e^{2i\phi(x=0)/\nu}$,
where the "2" is because two electrons are tunnelling.
Performing the average in (7) using the quadratic edge action (5) gives
$\delta = 2/\nu$.  Thus, for all odd integer $\nu^{-1}$, the tunnelling
amplitude $t$ flows to zero at low energies.
For $\nu=1$ this result is perhaps surprising, since one might have expected a
constant (energy-independent) tunnelling amplitude for a Fermi liquid edge
state.

At non-zero temperatures the RG flows are cut of by $T$, and one
obtains an effective temperature dependent tunnelling amplitude,
$t_{eff}(T) \sim t T^{2/\nu -1}$.  One expects the tunnelling conductance
through the junction to vary as $t_{eff}^2$, which gives the result

\begin{equation}
G(T) \sim t^2 T^{4/\nu -2}, ~~~~~~ \nu^{-1} = odd~integer~~~~.
\label{cond}
\end{equation}
This result can be verified directly by calculating the conductance via
a Kubo-type formula\cite{Kane} (see below).  One can also calculate
the non-linear current-voltage (I-V) curve throught the point
contact\cite{Kane}, and at $T=0$
one obtains $I \sim V^{4/\nu -1}$ for small $V$.

Equation (8) indicates that the conductance of a junction
separating a dissipationless superconductor
from a dissipationless quantum Hall fluid vanishes as $T \rightarrow 0$.
The point contact is insulating.  For the $\nu=1/3$ state, the
conductance vanishes with an enormous power, $G \sim T^{10}$.
For the integer state $\nu =1$ the power is smaller, with $G(T) \sim T^2$,
but even a vanishing conductance is surprising since the edge state is
a Fermi liquid in this case.

In order to understand the above result for $\nu=1$, it is helpful to calculate
explicitly the junction conductance using the non-interacting electron action
in
Eqn.(4).
The junction conductance is defined as

\begin{equation}
G=lim_{\omega \rightarrow 0} [ {1\over{\hbar \omega_n}} \int_{0}^{\beta}
d \tau e^{i\omega_n \tau} <T_\tau I(\tau)I(0) > ] \mid_{i\omega_n \rightarrow
\omega + i\epsilon}
\label{kubo}
\end{equation}
where the junction current operator is
\begin{equation}
I = 2eit [ \psi^\dagger (x=\xi) \psi^\dagger (x=0) \hat{c}(x=0) - h.c. ]  .
\label{curr}
\end{equation}
To leading (second) order in the tunnelling amplitude it is sufficient to
evaluate the correlation function in Eqn.(9) using the free fermion edge action
Eqn.(4).
Once again the superconducting pair field operator $\hat{c}$ can be
replaced by $\Delta$.
We thereby obtain a perturbative expression for the junction
conductance when $\nu=1$:
\begin{equation}
G= (4e^2/\pi v) (t\Delta)^2 \beta \int_{-\infty}^{\infty}
[1+cosh(\beta E)]^{-1} \rho(E)  .
\label{cond1}
\end{equation}
Here we have defined a "pair-tunnelling" density of states,
$\rho (E)=[1-cos(2\xi E/\hbar v)]/(\hbar v)$.  Evaluating this expression in
the
low temperature limit, $k_BT << \hbar v/\xi$, gives

\begin{equation}
G(T)= (4e^2/\hbar) (4\pi /3) (t\Delta \xi /\hbar v^2)^2 (k_B T)^2 .
\label{cond2}
\end{equation}
Notice that the $T^2$ dependence can be traced directly to the supression of
the
pair-tunnelling density of states, $\rho(E)$, which vanishes as $E^2$ for
$E<<\hbar v/\xi$.
This can also be seen in the expression for the tunnel current at finite
voltage, which at low temperatures is found to take the form
\begin{equation}
I \sim \int_{-\infty}^{\infty} dE \rho(E) [f(E-2eV)-f(E+2eV)] \sim V^3   ,
\label{iv}
\end{equation}
with $f(E)$ a Fermi function.
Physically, the suppression of the pair-tunnelling density of states can be
attributed
to the Pauli exclusion between the $\it{pair}$ of electrons.
After tunnelling one spin-up electron into the
edge state, tunnelling of the second electron is suppressed by
the Pauli exclusion principle up to a time $\xi /v$, at which point the first
electron
has been carried away a distance $\xi$ by the edge current.  This leads in turn
to a suppression in the pair-tunnelling density of states, $\rho (E)$, below
the energy
scale $\hbar v/\xi$.

It is amusing that this "Pauli blockade" of pair-tunnelling, effective
when $\nu=1$, can be circumvented when tunnelling into a $\nu=2$ state, which
has two edge channels.  Specifically, the pair of electrons can simultaneously
tunnel into the two different edge channels. This can be quantified as follows.
 Let $p$ denote the probability that
an electron in the pair will tunnel into the first ($j=1$) edge mode, and $1-p$
the
probability to tunnel into the second ($j=2$).  The electron
operators entering into the tunnelling Hamiltonian Eq.(3), can then be
expressed as $\psi=\surd{p}
\psi_1 + \surd(1-p) \psi_2$, where $\psi_j$ is the electron operator
in the $j^{th}$ edge state.  To evaluate
the junction conductance one needs the edge Greens functions, which follow
from the action in (4) and are given by

\begin{equation}
<T_\tau \psi_j(x,\tau) \psi^\dagger_j(0,0)> = {{e^{ik_jx}} \over { v_j \tau
+ ix }}  .
\label{green}
\end{equation}
Notice that we have included a phase factor into the above
Green's functions, with $k_j$ playing the role of an edge (Fermi-)momentum.
In general this edge momentum is gauge dependent, but the difference,
$k_{12}=k_1 - k_2$, is gauge invariant and determined by the
magnetic flux which penetrates between the two edge branches\cite{edge1}.
With $l$ denoting the spatial separation between the two branches
one has $k_{12} = eBl/\hbar$, with $B$ the magnetic field.

Using the above Greens functions it is straightforward to evaluate the
junction conductance for the case $\nu=2$, to leading order in the tunnelling
amplitude, $t$.  A straightforward calculation using Eqn.(9) gives the
conductance at zero temperature:

\begin{equation}
G(T=0)= (16e^2/\pi\hbar) (t\Delta / v)^2 p(1-p) [1-cos(k_{12}\xi)],~~~~~\nu=2 .
\label{cond3}
\end{equation}
Note that, in contrast to the $\nu=1$ case,
the tunnelling conductance for $\nu=2$ does $\it{not}$ vanish at $T=0$.  Pauli
exclusion is less effective here, with the two electrons tunnelling into
different edge states.  The
conductance is proportional to $p(1-p)$, and so vanishes when the tunnelling is
completely into one or the other of the two edge modes. Note, moreover,
that the edge momentum difference $k_{12}$ plays a crucial role
here.  Indeed, in the limit $k_{12} \rightarrow 0$, in which the edge states
sit atop one another, the tunnel conductance
vanishes.  In this limt one recovers the "Pauli
blockade".

In summary, I have introduced and analyzed a simple model for a tunnel
junction between a superconductor and a quantum Hall fluid.
The low temperature transport, dominated by pair-tunnelling, is suppressed
due to both Coulomb blockade effects and the Pauli exclusion principle.
The tunnelling conductance has been found to vanish as a power
law in temperature, which should be testable in
experiments on superconductor-2DEG samples.
Numerous interesting issues have not been addressed here, such as a.c.
transport,
resonant tunnelling and non-equilibrium noise at the superconductor/quantum
-Hall junction.

I would like to thank Paul Goldbardt for suggesting to me that Andreev
reflection might be modified in an interesting way in a Luttinger liquid.  I
have benefited from numerous clarifying conversations with David Morse.
This work was supported by the National Science Foundation under
grant No. PHY89-04035.

\newpage

FIGURE CAPTION

%\begin{itemize}
Schematic of tunnel junction from superconductor (s.c.) to a quantum Hall
fluid.  Cooper pairs tunnel into the edge at $x=0$.

\end{document}